\documentclass[prl,aps,twocolumn,amsmath,amssymb]{revtex4-1}
\pdfoutput=1
\usepackage{hyperref}
\usepackage{graphicx}
\usepackage{color}
\usepackage{multirow}
\usepackage{amsmath}

\usepackage{cleveref}

\newcommand{\be}{\begin{equation}}
\newcommand{\ee}{\end{equation}}
\newcommand{\bea}{\begin{eqnarray}}
\newcommand{\eea}{\end{eqnarray}}
\def\c#1{~\cite{#1}}
\def\f#1{Fig.~\ref{#1}}

\def\cot{CO$_2$}

\def\water{H$_2$O}

\def\eqq#1{Eq.~(\ref{#1})}
\def\eq#1{(\ref{#1})}
\def\av#1{\langle #1 \rangle}

\def\beq{\begin{equation}}
\def\eeq{\end{equation}}

\def\kt{k_{\rm B}T}

\begin{document}

\title{Cooperative gas adsorption without a phase transition in metal-organic frameworks}
\author{Joyjit Kund$\mbox{u}^{1,2}$}
\email{jkundu@lbl.gov}
\author{J{\"u}rgen F. Stilc$\mbox{k}^3$}
\author{Jung-Hoon Le$\mbox{e}^{1,4}$}
\author{Jeffrey B. Neato$\mbox{n}^{1,4,5}$}
\author{David Prendergas$\mbox{t}^1$}
\email{dgprendergast@lbl.gov}
\author{Stephen Whitela$\mbox{m}^1$}
\email{swhitelam@lbl.gov}
\affiliation{$^1$Molecular Foundry, Lawrence Berkeley National Laboratory, Berkeley, CA 94720, USA}
\affiliation{$^2$Department of Chemistry, Duke University, Durham, North Carolina 27708, USA}
\affiliation{$^3$Instituto de F\'{\i}sica and National Institute of Science and Technology for Complex Systems, Universidade Federal Fluminense, Av. Litor\^anea s/n, 24210-346 - Niter\'oi, RJ, Brazil }
\affiliation{$^4$Department of Physics, University of California, Berkeley, California 94720-7300, USA}
\affiliation{$^5$Kavli Energy Nanosciences Institute at Berkeley, Berkeley, California 94720, USA}

\date{\today}
\begin{abstract}
Cooperative adsorption of gases by porous frameworks permits more efficient uptake and removal than does the more usual non-cooperative (Langmuir-type) adsorption. Cooperativity, signaled by a step-like isotherm, is usually attributed to a phase transition of the framework. However, the class of metal-organic frameworks mmen-M$_2$(dobpdc) exhibit cooperative adsorption of \cot~but show no evidence of a phase transition. Here we show how cooperativity emerges in these frameworks in the absence of a phase transition. We use a combination of quantum and statistical mechanics to show that cooperativity results from a sharp but finite increase, with pressure, of the mean length of chains of \cot~molecules that polymerize within the framework. Our study provides microscopic understanding of the emergent features of cooperative binding, including the position, slope and height of the isotherm step, and indicates how to optimize gas storage and separation in these materials.
\end{abstract}

\maketitle

{\em Introduction --} The release of \cot~into the atmosphere due to the burning of fossil fuels causes climate change\c{pachauri2014climate}, and so it is important to develop technologies for \cot~capture and storage. Promising candidates in this regard are metal-organic frameworks (MOFs), porous crystalline materials with tunable molecular properties and large internal surface areas\c{yaghi2009,queen2014,long_chemrev,yaghi2013,yaghi1999,yaghi2003}. In equilibrium~\footnote{MOFs can also be productively used out of equilibrium: see e.g. Refs.\c{piero2013,kundu2016}}, most gas adsorption within MOFs can be described by Langmuir-type adsorption isotherms, in which the quantity of adsorbed gas varies gradually with pressure or temperature\c{long_chemrev,matzger2008,krishna_nat_comm,patric2013,hiro2009,yaghi_mm}. It is technologically more convenient, however, to have the quantity of adsorbed gas vary in an abrupt or step-like way with pressure and temperature. This phenomenon is known as cooperative adsorption, and is exhibited by a small handful of gas-framework combinations. These include CO adsorption in  Fe$_2$Cl$_2$(bbta)\c{long_Fe}, ${\rm CH}_4$ adsorption in Fe(bdp)\c{long2015}, and \cot~adsorption in diamine-grafted MOFs\c{david2015,chemsuschem}, in MIL-53\c{mill53,materials}, and in a bifunctional MOF\c{bifunc}. Cooperativity in most of these cases is attributed to a first-order phase transition~\c{cory_PNAS,coudert2008,coudert2011,coudert2013,smit2013,manos2016} or a dynamic rearrangement\c{seo_jacs,cory_PNAS} of the framework. Although, it has been established that \cot~molecules form ammonium carbamate chains at high pressures within the class of diamine-grafted MOFs $(\mbox{mmen-, en-, men-, or den-M}_2$(dobpdc), where M stands for the metal Mg, Mn, Fe, Co, or Zn), there exists no evidence of a phase transition or structural dynamism of the framework. Therefore, it is not clear why formation of such chains would lead to a step-like adsorption isotherm, or how to control it.

Here we show how this cooperativity emerges in the absence of an underlying phase transition. Experimental studies and quantum mechanical density-functional theory (DFT) calculations had previously revealed that, at low partial pressure of \cot,~the gas molecules are adsorbed as single molecules or as carbamic acid pairs\c{smit2015}. At high partial pressure, by contrast, \cot~undergoes chemisorption, by forming one-dimensional ammonium carbamate chains that run down the channels of the MOF, along the $c$-axis\c{david2015}. The statistical mechanics of one-dimensional structures\c{goldenfeld} indicates that chain formation cannot be accompanied by a phase transition: finite-temperature phase transitions in one dimension require long-range interactions, and there are no indications of long-range interactions in the system (either direct or mediated by the framework). By mapping \cot~adsorption in $\mbox{mmen-M}_2$(dobpdc) to an exactly solvable statistical mechanical model, parameterized by our DFT calculations, we show that the mean chain length of \cot~within the MOF-pores undergoes a sharp change with pressure, leading to cooperativity in the absence of an underlying phase transition. Amine-functionalized MOFs have emerged as one of the best framework types for \cot~capture and separation because, unusually for MOFs, they capture \cot~selectively in the presence of water\c{long_mixture,david2015,chemsuschem}. Our results provide a microscopic understanding of cooperativity in these MOFs, and reveal strategies for its control. In what follows we describe our calculations and their implication for optimizing \cot~capture in experiments.
\begin{figure*}
\includegraphics[width=0.80\linewidth]{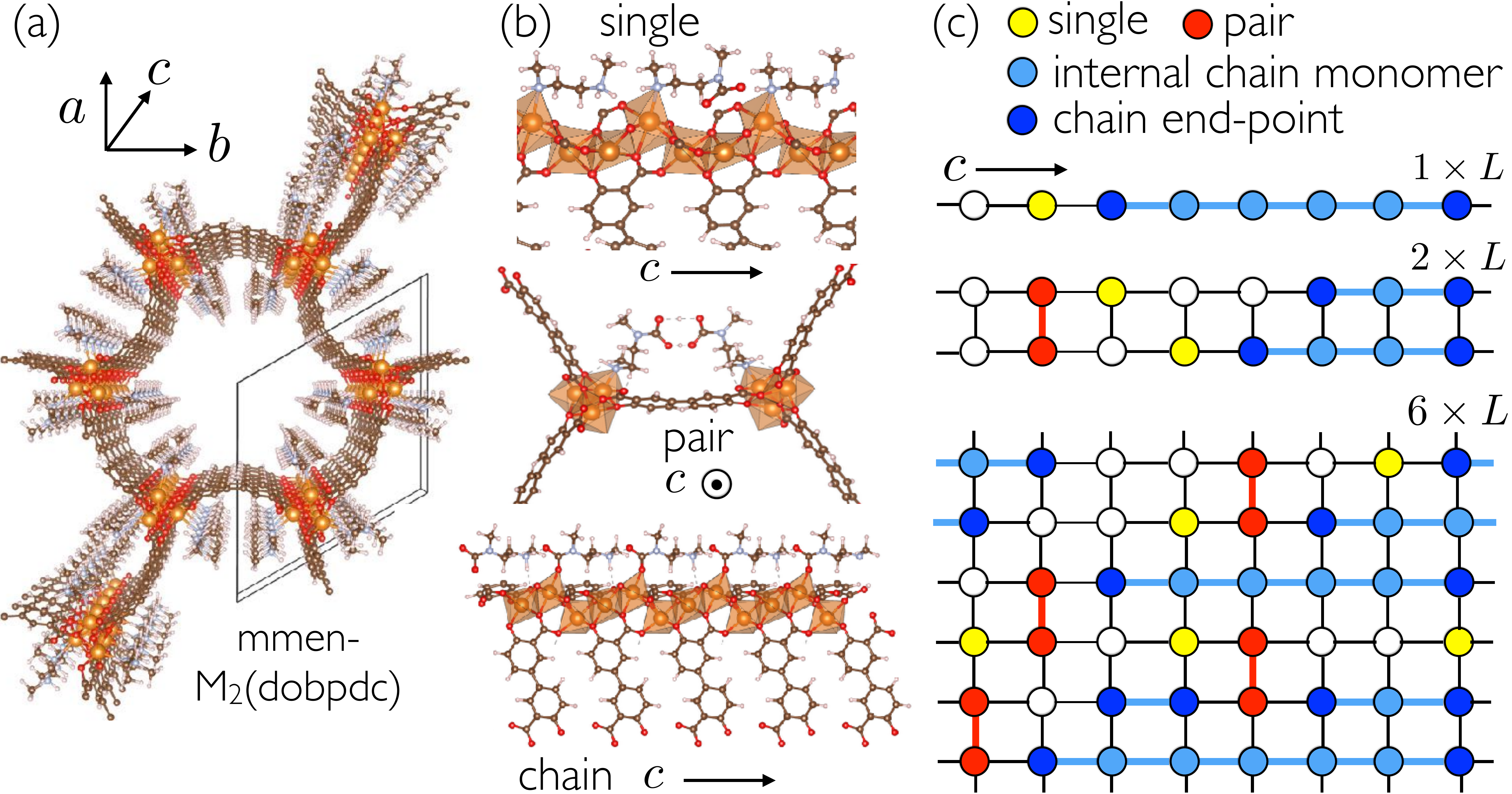}
\caption{(a) Hexagonal channel of mmen-M$_2$(dobdpc) (where M stands for the metal Mg, Mn, Fe, Co, Zn or Ni). (b) The three possible conformations of adsorbed \cot~within this class of MOF. (c) Our statistical mechanical models of mmen-M$_2$(dobdpc), in example configurations.}
\label{fig01}
\end{figure*}

{\em Model --} We start by considering binding geometries and affinities of \cot~within mmen-$\mbox{M}_2$(dobpdc). \cot~can bind within this class of MOFs as 1) a single molecule, 2) a bound (carbamic acid) pair, or 3) as part of a polymerized (ammonium carbamate) chain of molecules involving the ligands through its insertion at the metal sites\c{david2015}. Pairs form in the $ab$-plane\c{smit2015}; see \f{fig01}(a,b). By contrast, chains are formed parallel to the $c$-axis, along any of 6 lanes around the periphery of the MOF channel, but usually do not interact in the $ab$-plane\c{long_2017}. For mmen-MOF built from the metals Mg, Mn, Fe, Co, and Zn, experiments and DFT calculations show that molecules in the chain conformation are lower in energy than molecules in the single- and pair conformations\c{david2015,smit2015} (see Table~I in the SI). Our DFT calculations (see SI Sections~S1 and~S2) also indicate that \cot~molecules at the end of a chain are higher in energy than those in the interior of a chain. Entropically, by contrast, the chain conformation is less favorable than the other two conformations, because a \cot~molecule within a chain can access less free volume than can a molecule in the single-molecule or pair conformation. 

\begin{figure*}
\includegraphics[width=\linewidth]{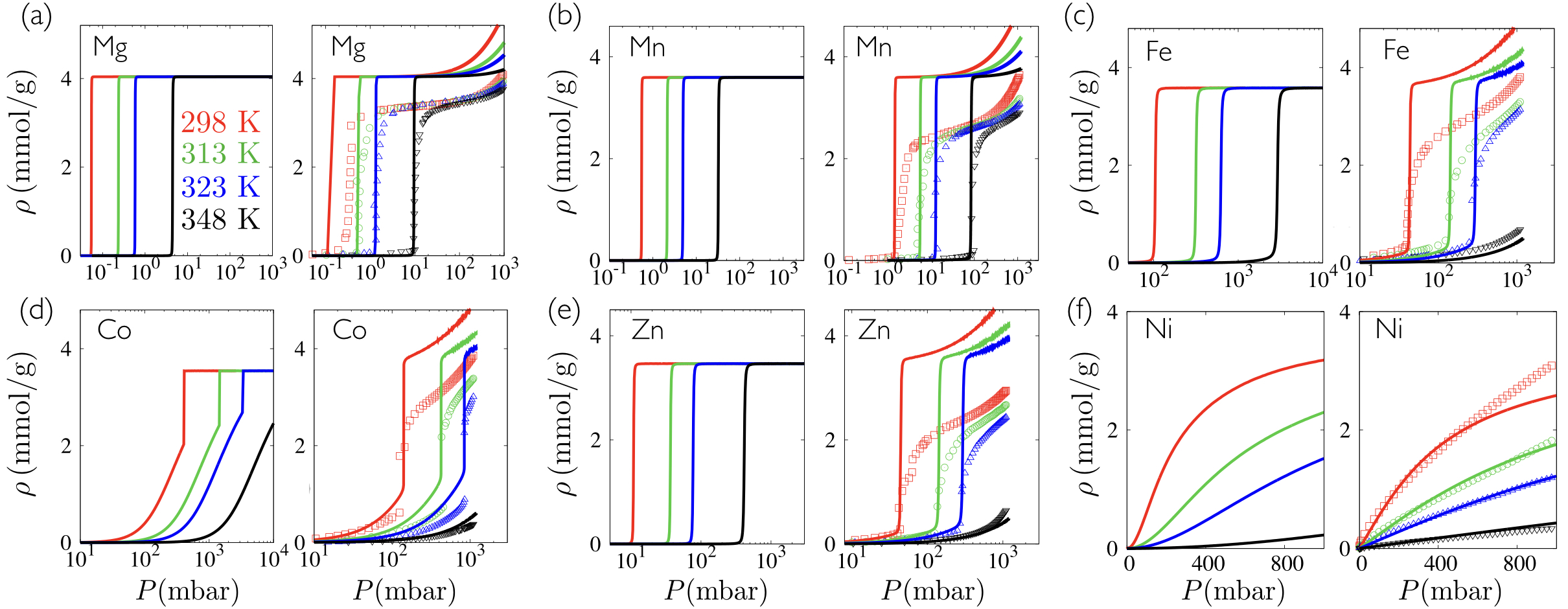}
\caption{Isotherms calculated from our model of mmen-$\mbox{M}_2$(dobpdc) (lines) versus experimental data (symbols). In left-hand panels the model is parameterized using quantum mechanical data, and captures the sharp isotherm, and trend in step pressure with temperature, seen in experiment\c{david2015}. In right-hand panels we use the experimental binding enthalpy values, where available, to identify the model parameter $V_{\rm int}$ that gives the best match with experimental data; see SI Table~II (for Ni, we vary the parameter $E_{\rm d}$ to obtain the best-fit; see SI Sec.~S5). For Mg and Mn we can ignore the pair conformation, which is energetically disfavored, and use the 1-lane model. The other panels are derived from the 6-lane model, which accounts for pair binding. Here the model parameters $E_1$, $E_{\rm d}$, and $E_{\rm int}$ are taken from Table~I in the SI; $V_1=500 \AA$ and $V_{\rm d}=75 \AA$.}
\label{isotherm}
\end{figure*}

The thermodynamics of this system can be described by the equilibrium polymerization model\c{wheeler1983}, sketched in \f{fig01}(c). This is a lattice model, extended in one dimension (corresponding to the $c$-axis of the MOF). Lattice sites can be vacant, occupied by a single particle (a \cot~molecule), or occupied by a particle that is a member of a pair or a chain of particles. We further distinguish chain end sites from chain interior sites. In some versions of mmen-$\mbox{M}_2$(dobpdc), e.g. where M is Mg or Mn, the bound-pair binding affinity is small enough, relative to the chain, that it can be ignored\c{smit2015} (see Table~I in the SI). In these cases it is sufficient to consider a $1$-lane model, which represents one of the six independent lanes running along the $c$-axis. In the presence of the bound-pair conformation we need to allow finite extent in the $ab$-direction. A 6-lane model is then required to describe all possible \cot~conformations within the 6-lane MOF channel. We have also considered a 2-lane model, because the distance between the lanes in some versions of mmen-$\mbox{M}_2$(dobpdc) is such that the framework is best described as 3 independent 2-lane structures (i.e. \cot-pairs can only bridge alternate pairs of lanes). We have solved the 1-, 2- and 6-lane models exactly. The 1-lane model captures the basic physics of cooperative binding in all experiments we consider. The 2-lane and 6-lane models capture, in addition, fine features of adsorption isotherms seen in MOFs in which pair-binding is significant (see SI Sec.~S3 for details).

{\em Model solution -- } We start with the 1-lane model. Let the statistical weights for a single bound molecule, a molecule internal to a chain, and a molecule at either end-point of a chain be $g_1 W_1$, $g_{\rm int} W_{\rm int}$, and $g_{\rm end} W_{\rm end}$, respectively. Here $g_\alpha=V_\alpha \Lambda^{-3} q_{\rm inter}$ ($\alpha=\{1,{\rm int},{\rm end}\}$). The factor $V_\alpha \Lambda^{-3}$ arises from the configurational partition sum and is related to the translational entropy of the adsorbate; $\Lambda$ is the de Broglie wavelength; and $V_\alpha$ is the free volume accessible to the adsorbate in the conformation $\alpha$\c{cory_PNAS}. The factor $q_{\rm inter}$ is the partition sum of \cot~due to its internal degrees of freedom~\footnote{$q_{\rm inter}=q_{\rm rot} q_{\rm vib}=\frac{8 \pi^2 I \kt}{\sigma h^2} \prod_{j=1}^4\frac{\exp(-\theta_{\rm{vib},j}/2T)}{1-\exp(-\theta_{\rm{vib},j}/2T)}$, where $q_{\rm rot}$ and $q_{\rm vib}$ are the rotational and vibrational partition functions of a \cot~molecule, $I$ is the moment of inertia and $\sigma$ $(=2)$ is the symmetry factor of \cot. ($\theta_{\rm{vib},j}=h \nu_j /k_{\rm B}$; $\nu_j$ corresponds to the frequency of $j$-th normal mode of vibration)}. These statistical weights can be related to the energy of a particle in conformation $\alpha$ via $W_\alpha=\exp[\beta(\mu-E_\alpha)]$, where $\beta \equiv 1/(\kt)$, and $\mu$ is the chemical potential, set by the pressure $P$ of \cot~in the bulk. We convert $\mu$ to $P$ using the ideal gas relation for a linear triatomic molecule, $e^{\beta \mu}=\beta P \Lambda^3/q_{\rm inter}$~\footnote{This approximation is reasonable because cooperative behavior in experiment is observed at pressures of 1 bar or below, where \cot~behaves as an ideal gas; using the Peng-Robinson equation of state for \cot~we checked that the fugacity coefficient of \cot~at 1 bar is $\approx 0.99$}. To simplify notation we define $K_\alpha \equiv g_\alpha W_\alpha$. We then have $K_\alpha=\beta P V_\alpha e^{-\beta E_{\alpha}}$~\footnote{$K_\alpha=g_\alpha W_\alpha=V_\alpha \Lambda^{-3} q_{\rm inter} \exp(\beta \mu) \exp (-\beta E_\alpha)=V_\alpha \Lambda^{-3} q_{\rm inter} (\beta P \Lambda^3 q^{-1}_{\rm inter}) \exp (-\beta E_\alpha)=\beta P V_\alpha e^{-\beta E_{\alpha}}$}. We set $V_1=500~\AA^3$, $V_{\rm int}=V_{\rm end}=11\AA^3$ using simple geometric arguments  (see SI Sec.~S4): the single bound \cot~molecule has orientational entropy associated with the corresponding diamine, while \cot~in the chain conformation is almost frozen. 

The grand partition function is\bea
\mathcal{Z}=\sum_{\{n_1,n_{\rm int},n_{\rm end}\}} K_1^{n_1} K_{\rm end}^{n_{\rm end}} K_{\rm int}^{n_{\rm int}} \Gamma (n_1,n_{\rm int},n_{\rm end}),
\eea
where $\Gamma$
is the number of ways of arranging $n_1$ single \cot~molecules, $n_{\rm int}$ internal chain molecules and $n_{\rm end}$ chain end-points on a $1d$ lattice with $N$ sites (see Eq.~S4 in the SI). To solve this model one can introduce the restricted partition functions $Z^{\rm u}_N$ and $Z^{\rm p}_N$ for a system of $N$ lattice sites that possesses an external edge (connected to a notional $(N+1)^{\rm th}$ site)\c{wheeler1983}. This external edge is specified to be, respectively, unpolymerized (not within a chain conformation) or polymerized (internal to a chain). $Z^{\rm u}_{N+1}$ and $Z^{\rm p}_{N+1}$ can be expressed in terms of $Z^{\rm u}_N$ and $Z^{\rm p}_N$ as $\overline{Z}_{N+1}=T \overline{Z}_N$, where $\overline{Z}=\left( \begin{array}{c} Z^{\rm u}\\ Z^{\rm p} \end{array} \right)$ and $T$ is the transfer matrix,
\begin{equation}
T=\begin{pmatrix} 1+ K_1 & K_{\rm end} \\ K_{\rm end} &  K_{\rm int} \end{pmatrix}.
\end{equation}
Using the boundary conditions $Z^{\rm u}_1=1+K_1$ and $Z^{\rm p}_1=K_{\rm end} $ we can write $Z^{\rm u}_{N}=\begin{pmatrix} 1 & 0\end{pmatrix} T^N \begin{pmatrix} 1 \\ 0\end{pmatrix}$. The partition function can then be expressed as\c{wheeler1983}
\begin{equation}
Z^{\rm u}_N=\frac{\lambda_+^N (1+ K_1-\lambda_-)+\lambda^N_1 (\lambda_+-1-  K_1)}{\lambda_+-\lambda_-},
\end{equation}
in terms of the eigenvalues $\lambda_\pm$ of $T$, where
\beq
\label{ev}
2 \lambda_{\pm}=1+K_1+K_{\rm int} \pm \sqrt{(1+K_1- K_{\rm int})^2+4 K_{\rm end}^2}.
\eeq
In the thermodynamic limit the free energy is $f=-\kt \ln \lambda_+$ which has a singularity (and so admits a phase transition) only in the experimentally inaccessible limit in which chain end-points are energetically infinitely unfavorable ($K_{\rm end}=0$, with $1+K_1= K_{\rm int}$). For experimental parameters the free energy is analytic, and so no phase transition occurs.
\begin{figure*}[t]
\includegraphics[width=\linewidth]{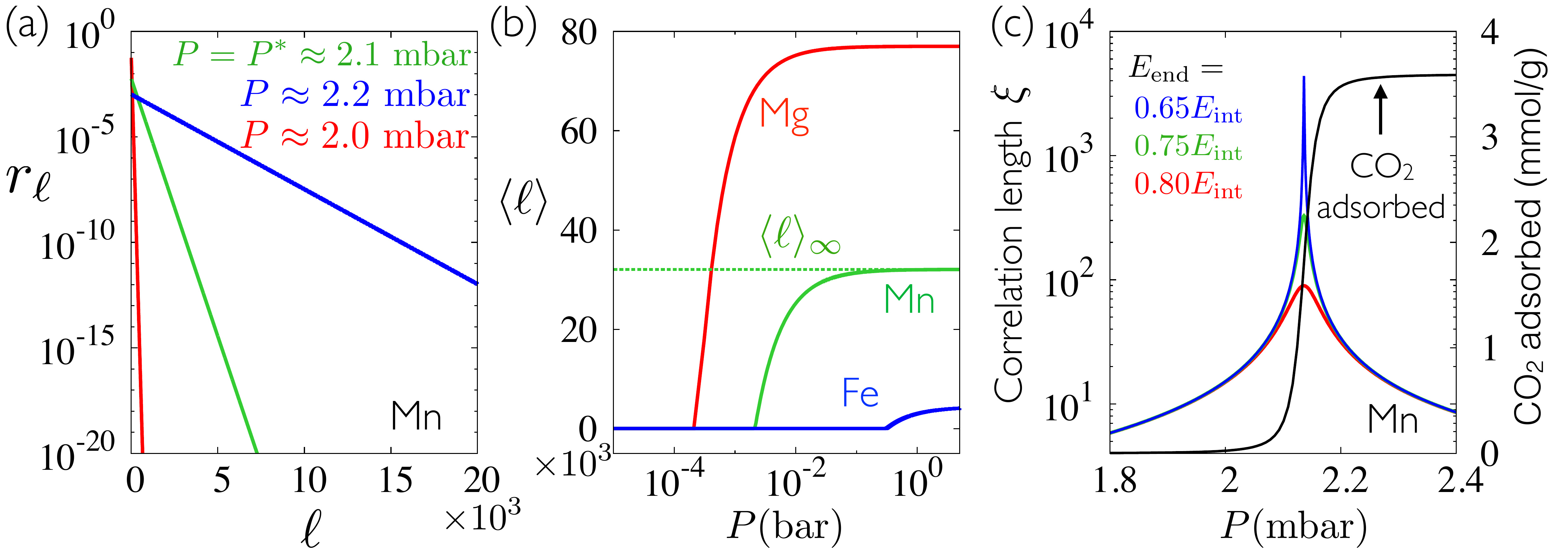}
\caption{(a) Chain-length distribution $r_\ell$ at different pressures for Mn at $313$ K. (b) Mean chain length $\av{\ell}$ as a function of pressure for different metals at $313$ K. Here length is expressed in units of $7 \AA$, the unit-cell spacing of the crystal structure along the $c$-axis. (c) Bond-bond correlation length $\xi$ (in unit of $c$-spacing) as a function of pressure $P$ for Mn at $313~\mbox{K}$ (colored lines), together with the adoption isotherm. Note that $\xi$ grows exponentially with the energy cost for chain termini, $E_{\rm end}$, for a given internal chain-monomer energy $E_{\rm int}$.}
\label{len_dist}
\end{figure*}

{\em Model-experiment comparison --} Despite the absence of a phase transition, the isotherm of adsorbed \cot~versus pressure displays a sharp step (when $K_{\rm int}> K_1,K_{\rm end}$) similar to those seen in experiment; see \f{isotherm}. To convert lattice-site occupancies  $\rho=P(\partial f/\partial P)$ to experimental units we multiply our calculated density by the theoretical maximum uptake capacity ($q_{\rm M}$) of the MOF for each M. The values of $q_{\rm M}$ are listed in Table~III in the SI. For each metal, the isotherms in the left-hand panels in \f{isotherm} are generated from first principles, using binding energies and enthalpies obtained by DFT calculations (Table~I in the SI), while the right-hand panels contain experimental data. The comparison shows that a combination of quantum and statistical mechanics, with no experimental input, can reproduce the sharp step seen in experimental isotherms, and can capture the trend in step-pressure with temperature. 

For the metals Mg and Mn, considered in \f{isotherm}(a,b), we use the 1-lane model, because pair-binding is energetically disfavored. For the other metals we use the 6-lane model (detailed in SI Sec.~S3), because the bound-pair conformation, characterized by binding energy $E_{\rm d}$ and free volume $V_{\rm d}$, is free-energetically significant (Table~I in the SI). Note that the isotherm for Nickel (panel (f)) has no sharp step, because chain polymerization does not occur, at least, at those pressures.

The basic physics of adsorption in all cases is captured by the simple considerations described above. In the right-hand panels of \f{isotherm} we show that additional fine features of binding, such as the rise of isotherms before and after the step, can be captured by including within the model two additional physical ingredients, namely the existence of secondary binding sites, and of a different mode of monomer binding. Details of these calculations are given in SI Sec.~S5. Thus the model can provide insight into both the basic physics and the fine details of cooperative binding (e.g. the occupancy of different species as a function of pressure, measurable in NMR experiments; see Fig. S1).

{\em Origin of cooperative binding --} The microscopic origin of the step in adsorption isotherms is a sudden but finite increase, with pressure, of the mean length of chains of \cot. To demonstrate this fact we calculate the chain-length distribution exactly, using the transfer matrix technique\c{jurgen_physica}, as described in SI Sec.~S6. The fraction of chains of length $\ell$, $r_\ell$, can be expressed in terms of densities of chain-internal monomers ($\rho_{\rm int}$) and end-points $(\rho_{\rm end})$, as
\beq
\label{dist}
r_\ell = \frac{\rho_{\rm end}}{\rho_{\rm end}+2 \rho_{\rm int}} \exp[-(\ell-2) /\ell_0],
\eeq
where $\ell_0 \equiv -1/\ln \left[2 \rho_{\rm int}/(\rho_{\rm end}+2 \rho_{\rm int})\right]$; $\rho_{\rm end}=2 K_{\rm end}^2 \omega (1-\omega K_{\rm int})/D$; $\rho_{\rm int}=K_{\rm end}^2 \omega (\omega K_{\rm int})/D$; $D=(1+K_1)(1-\omega K_{\rm int})^2+K_{\rm end}^2 \omega (2-\omega K_{\rm int})$; and $\omega \equiv 1/\lambda_+$. The average chain length is $\av{\ell}=2+\rho_{\rm end} \ell_0^2 /({\rho_{\rm end}+2 \rho_{\rm int}})$. 

From \eqq{dist} we see that the chain-length distribution for the $1$-lane model decays exponentially at all pressures, including at the step pressure $P^{\ast}$ (which satisfies $d^2\rho/dP^2|_{P^{\ast}}=0$). In \f{len_dist}(a) we plot the chain-length distribution $r_\ell$ for Mn. In panel (b) we plot the mean chain length $\av{\ell}$, as a function of pressure, for different metals at $313$ K. \cot~molecules undergo polymerization beyond a threshold pressure, leading to a sharp (but finite) increase of the mean chain length. This sharp increase results in the step-like feature of the isotherm (Langmuir-type behavior is recovered when $K_1 \gtrsim K_{\rm int}$). The rise is rather gradual when chain-end points are energetically equivalent to internal points (see SI Fig. S5). In the infinite-pressure limit the mean chain length tends to a finite value $\av{\ell}_{\infty}$ (given by Eq.~(S15) in the SI). For Mg and Mn at 313 K, for instance, $\av{\ell}_\infty \approx 53~\mu {\rm m}$ and $22~\mu {\rm m}$, respectively (the typical grain size in experiments is $\sim 10~\mu {\rm m}$\c{david2015}).

In \f{len_dist}(c) we show that the bond-bond correlation length (the distance over which fluctuations of bond occupancies are correlated) displays a (non-diverging) maximum at the step position (see SI Sec.~S7). The behavior shown \f{len_dist} looks superficially like a phase transition, but it is not: both the mean length of chains and the bond-bond correlation length remain finite.

{\em Conclusions --} We have shown that cooperative \cot~adsorption in the class of diamine-grafted metal-organic frameworks arises from an abrupt polymerization of \cot~molecules into long chains within the channels of the framework in the absence of a phase transition. Our calculations using a classic model of statistical mechanics\c{wheeler1983}, parameterized by quantum mechanical calculations, provide microscopic understanding of each feature of the cooperative isotherm, and so indicate how to alter these features for experimental convenience. For instance, the adsorption isotherm can be made more abrupt by increasing the penalty for chain end-points (see Fig. S5). In addition, understanding of cooperativity in these systems suggests ways of inducing cooperativity in gas-framework combinations in which it is absent (e.g. mmen-Ni$_2$(dobpdc)), by e.g. introducing additional binding agents (see SI Sec.~S5 and Fig.~S6).

\textit{Acknowledgements -- } We thank Rebecca Siegelman, M\'{a}rio J. de Oliveira, and Alexander C. Forse for discussions and comments on the manuscript. This work was done as part of a User project at the Molecular Foundry at Lawrence Berkeley National Laboratory, supported by the Office of Science, Office of Basic Energy Sciences, of the U.S. Department of Energy under Contract No. DE-AC02--05CH11231. JK and JHL were supported by the Center for Gas Separations Relevant to Clean Energy Technologies, an Energy Frontier Research Center funded by the U.S. Department of Energy, Office of Science, Basic Energy Sciences under award DE-SC0001015. JN, DGP, and SW were partially supported by the same Center. JFS was partially supported by CNPq. This research used resources of the National Energy Research Scientific Computing Center, a DOE Office of Science User Facility supported by the Office of Science of the U.S. Department of Energy under Contract No. DE-AC02-05CH11231, and used the Savio computational cluster provided by the Berkeley Research Computing program at the University of California, Berkeley (supported by the UC Berkeley Chancellor, Vice Chancellor for Research, and Chief Information Officer).
%

\onecolumngrid
\clearpage

~\\
\begin{center}
{\noindent\Large{Supplemental Information}}\\
\noindent \Large for\\
{\noindent \Large ``Cooperative gas adsorption without a phase transition in a metal-organic framework''}\\
~\newline
{\noindent \normalsize Joyjit Kund$\mbox{u}^{1,2}$, J{\"u}rgen F. Stilc$\mbox{k}^3$, Jung-Hoon Le$\mbox{e}^{1,4}$, Jeffrey B. Neato$\mbox{n}^{1,4,5}$, David Prendergas$\mbox{t}^1$ \& Stephen Whitela$\mbox{m}^1$}\\
{ \small
~\\
\noindent $^1$Molecular Foundry, Lawrence Berkeley National Laboratory, Berkeley, CA 94720, USA\\
$^2$Department of Chemistry, Duke University, Durham, North Carolina 27708, USA\\
$^3$Instituto de F\'{\i}sica and National Institute of Science and Technology for Complex Systems, Universidade Federal Fluminense, Av. Litor\^anea s/n, 24210-346 - Niter\'oi, RJ, Brazil\\
$^4$Department of Physics, University of California, Berkeley, California 94720-7300, USA\\
$^5$Kavli Energy Nanosciences Institute at Berkeley, Berkeley, California 94720, USA}


\end{center}\renewcommand{\theequation}{S\arabic{equation}}
\renewcommand{\thefigure}{S\arabic{figure}}
\renewcommand{\thesection}{S\arabic{section}}

\setcounter{equation}{0}
\setcounter{section}{0}
\setcounter{figure}{0}

\setlength{\parskip}{0.25cm}%
\setlength{\parindent}{0pt}%

\section{Density-Functional Theory calculations}
\label{dft}
In order to compute binding energies of \cot~gas molecules, we optimize mmen-M$_2$(dobpdc) MOFs without \cot~molecules (E$_{\rm{mmen-MOF}}$), 
CO$_2$ in the gas phase (E$_{\rm{CO}_2}$) within a $15\AA \times 15 \AA \times 15 \AA$  cubic supercell, and 
mmen-M$_2$(dobpdc) MOFs with CO$_2$ molecules (E$_{\rm{CO_2-mmen-MOF}}$) using vdW-corrected DFT. The binding energies (E$_{\rm B}$) are obtained via the difference

\begin{equation}
\label{eq:1}
E_{\rm B} = E_{\rm{CO_2-mmen-MOF}} - (E_{\rm{mmen-MOF}} + E_{\rm{CO_2}}).
\end{equation}

We also consider zero-point energy (ZPE) and thermal energy (TE) corrections to compare computed binding energies with experimentally determined CO$_2$ heats of adsorption, following a previous DFT study\c{kyuho2015}. We calculate vibrational frequencies of bound mmen and  CO$_2$-mmen in the framework and free mmen and  CO$_2$-mmen molecules within a $15\AA \times 15 \AA \times 15 \AA$ cubic supercell. Here we assume that phonon mode changes of the framework are small relative to those in molecular modes. All ZPE and TE corrections are obtained at 298 K.

We have estimated the \cot~binding energies for chains of different lengths: (i) two, (ii) three, and (iii) four in the $1\times 1\times 4$ supercell of mmen-$\mbox{Mg}_2$(dobpdc). In the case of length four, it is very close to a fully-occupied chain (periodic) since the calculations are performed in the $1\times 1\times 4$ supercell. In fact the \cot~ binding energy ($-65$ kJ/mol) in a chain of length four is about $10$ kJ/mol smaller than that of the unit-cell ($-75$ kJ/mol; without ZPE and TE corrections). This is because we fully relaxed the volume of the unit-cell while we did not relax the supercell when we computed the \cot~binding energies within short chains. In addition, we merely consider one channel of mmen ligands in the $1\times 1\times 4$ supercell of mmen-$\mbox{Mg}_2$(dobpdc). If we consider other channels of mmen ligands and relax the volume we get the same value as that of the unit-cell. As listed in Table~\ref{fig_table}, the average \cot~binding affinity increases with the chain length. We also compute the binding energy of a single bound \cot~within mmen-$\mbox{Mg}_2$(dobpdc), as shown in Table~\ref{fig_table} a. If the \cot~molecule is not directly bound to the metal, its binding energy should not depend on the metal-type. 

To quantitatively understand the cooperative \cot~capture mechanism, we perform ab-initio density functional theory (DFT) calculations within the generalized-gradient approximation (GGA) of Perdew, Burke, and Ernzerhof (PBE)\c{perdew1996}. We use a plane-wave basis and projector augmented wave (PAW)\c{blochl1994,kresse1999} pseudopotentials with the Vienna ab initio Simulation Package (VASP)\c{kresse1993,kresse1996,kresse_1996,hafner1994}. To assess the effect of the van der Waals (vdW) interaction on binding energies, we perform structural relaxations with corrections (vdW-DF2) for the vdW dispersion interaction as implemented in VASP\c{lee2010}. For all unit-cell calculations, Brillouin zone integrations are approximated using the $\Gamma$-point only, and we truncate the plane-wave basis using a 600 eV kinetic energy cut off. We explicitly treat $2$ valence electrons for each Mg ($3s^2$), 7 for Mn ($3d^5 4s^2$), 8 for Fe ($3d^6 4s^2$), 9 for Co ($3d^7 4s^2$), 12 for Zn ($3d^{10} 4s^2$), $6$ for O ($2s^2 2p^4$), $5$ for N ($2s^2 2p^3$), $4$ for C ($2s^2 2p^2$), and $1$ for H($1s^1$). To compute the \cot~binding energies for chains of different lengths, we use a plane-wave cut off of 500 eV and adopt a $1\times 1\times 4$ supercell of mmen-$\mbox{Mg}_2$(dobpdc). We only consider one channel of mmen ligands and relaxed the ions until the forces on them are less than $0.04~ \mbox{eV/\AA}$ while fixing the lattice parameters. The lattice parameters of the supercell are obtained from the fully-relaxed mmen-$\mbox{Mg}_2$(dobpdc) unit-cell.

\section{Binding energies and other model-parameterization data}
\label{data}

\begin{table}[h]
\begin{minipage}[b]{\linewidth}\centering
\centering
\begin{tabular}{c} 
(a) Single bound \cot~(this work): -22.6
\end{tabular}
\label{tab:PPer}
\end{minipage}
\\ \vspace{0.25 cm}
(b) Pair\c{smit2015}
\\ \vspace{0.10 cm}
\begin{minipage}[b]{\linewidth}\centering
\centering
\begin{tabular}{|c|c|c|c|c|c|c|} 
\hline 
  M& Mg & Mn & Fe & Co & Zn & Ni \\
\hline \hline
DFT & $-45.8$ & $-42.5$ & $-43.1$ & $-46.5$ & $-42.6$ & $-47.2$ \\
\hline 
\end{tabular}
\label{tab:PPer}
\end{minipage}
\\ \vspace{0.25 cm}
(c) Interior of a chain \\ \vspace{0.10 cm}
\begin{minipage}[b]{\linewidth}\centering
\centering
\begin{tabular}{|c|c|c|c|c|c|c|} 
\hline 
  M& Mg & Mn & Fe & Co & Zn & Ni \\
\hline \hline
Experiment\c{david2015}& $-71.0$ & $-67.0$ & $-58.0$ & $-52.0$ & $-57.0$ & $ - $ \\
\hline
DFT\c{smit2015}& $-69.4$ & $-66.8$ & $-57.7$ & $-50.8$ & $-50.8$ & $-46.4$ \\
\hline
DFT (this work)& $-73.2$ & $-67.5$ & $-55.5$ & $-52.4$ & $-60.1$ & $ - $ \\
\hline
\end{tabular}
\end{minipage}
\\ \vspace{0.25 cm}
(d) Chain with endpoints (this work) \\ \vspace{0.10 cm}
\begin{minipage}[b]{\linewidth}\centering
\centering
\begin{tabular}{|c|c|c|c|} 
\hline 
Chain length & $2$ & $3$ & $4$ \\
\hline \hline
Binding energy& $-35.0$ & $-44.6$ & $-65.1$ \\
\hline
\end{tabular}
\end{minipage}
\caption{Binding affinity (in kJ/mol) per \cot~molecule (a) of a single molecule bound at the free end of an mmen ligand, (b) of a carbamic acid pair, (c) in the interior of a chain, and (d) for short chains in case of mmen-Mg$_2$(dobpdc) (including chain end-points, e.g. a chain of length four has two end monomers and two internal monomers). By comparing (c) and (d), we set the binding enthalpy of \cot~molecules at the end of a chain as $E_{\rm end} \approx 0.8 E_{\rm int}$. The question whether chain conformation in mmen-Ni$_2$(dobpdc) is stable requires further investigation. Zero-point energy (ZPE) and thermal energy (TE) corrections of \cot-mmen and mmen ligands are considered (for our measurements) in case of chain-interior sites. All ZPE and TE values are obtained at 298 K.}
\label{fig_table}
\end{table}

\begin{center}
\begin{table}[h]
\begin{tabular}{|c|c|c|c|c|c|c|}
\hline
 M&Mg &  Mn & Fe & Co & Zn & Ni\\
\hline
\hline
Best fit & 11.0& 5.1 & 6.8 &27.4 &9.8 & 11.0 \\
\hline
\end{tabular}
\caption{Best-fit value of $V_{\rm int}$ ($\AA^3$) for different metal  types.}
\label{table02}
\end{table}
\end{center}

\begin{figure}
\includegraphics[width=0.65\columnwidth]{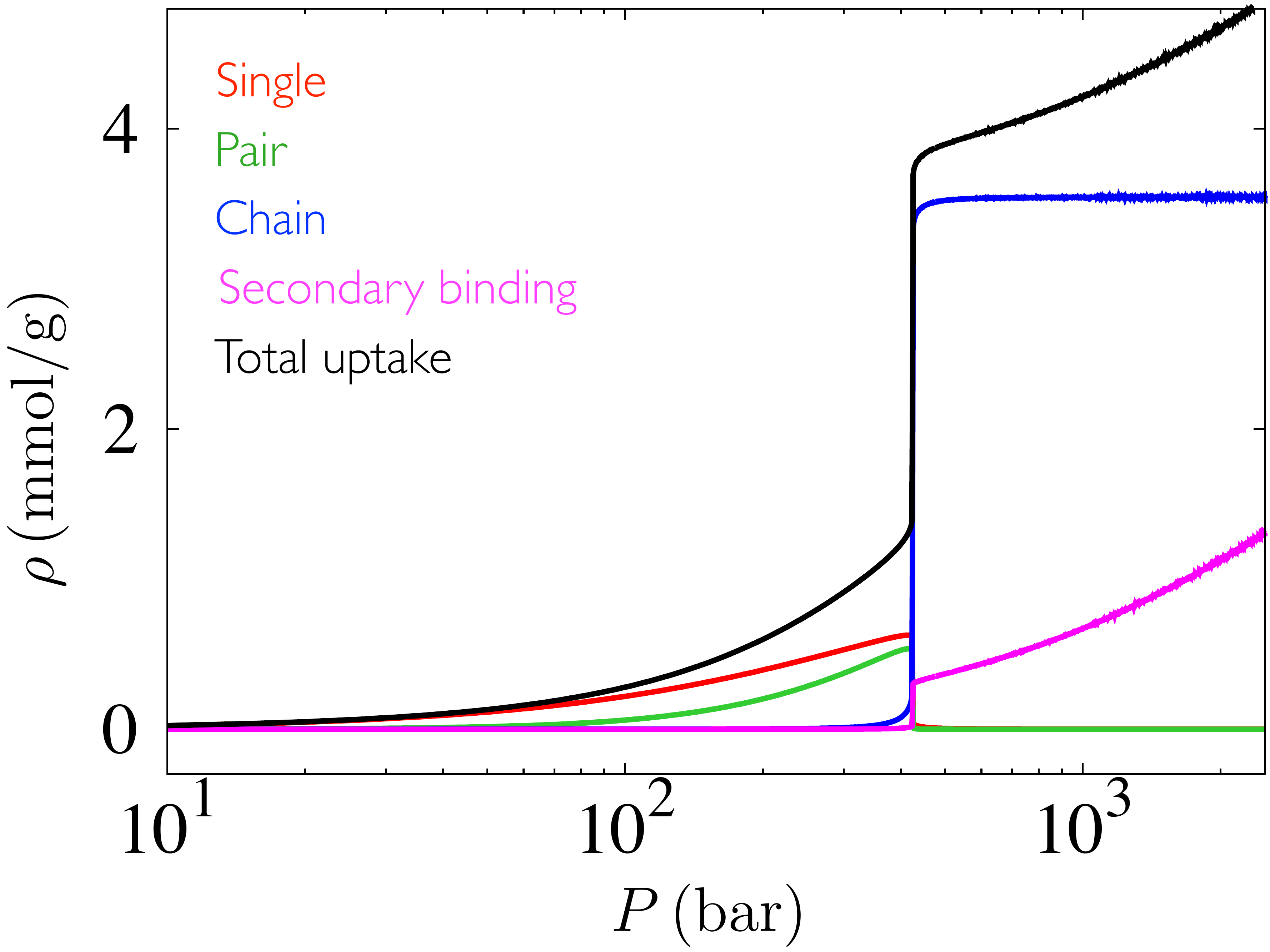}
\caption{Occupancy of different species as a function of pressure for mmen-$\mbox{Co}_2$(dobpdc) at $313$ K derived from the 6-lane model. Here, we consider adsorption at secondary binding sites, and use experimental binding enthalpy for the chain conformation along with the best-fit value of $V_{\rm int}$ (see Sec.~\ref{details}): $E_1=-40.5$ kJ/mol, $E_{\rm d}=-46.5$ kJ/mol, $E_{\rm int}=52.0$ kJ/mol, and $V_{\rm int}=27.4~\AA^3$.}
\label{comp}
\end{figure}
\begin{figure}
\includegraphics[width=0.42\columnwidth]{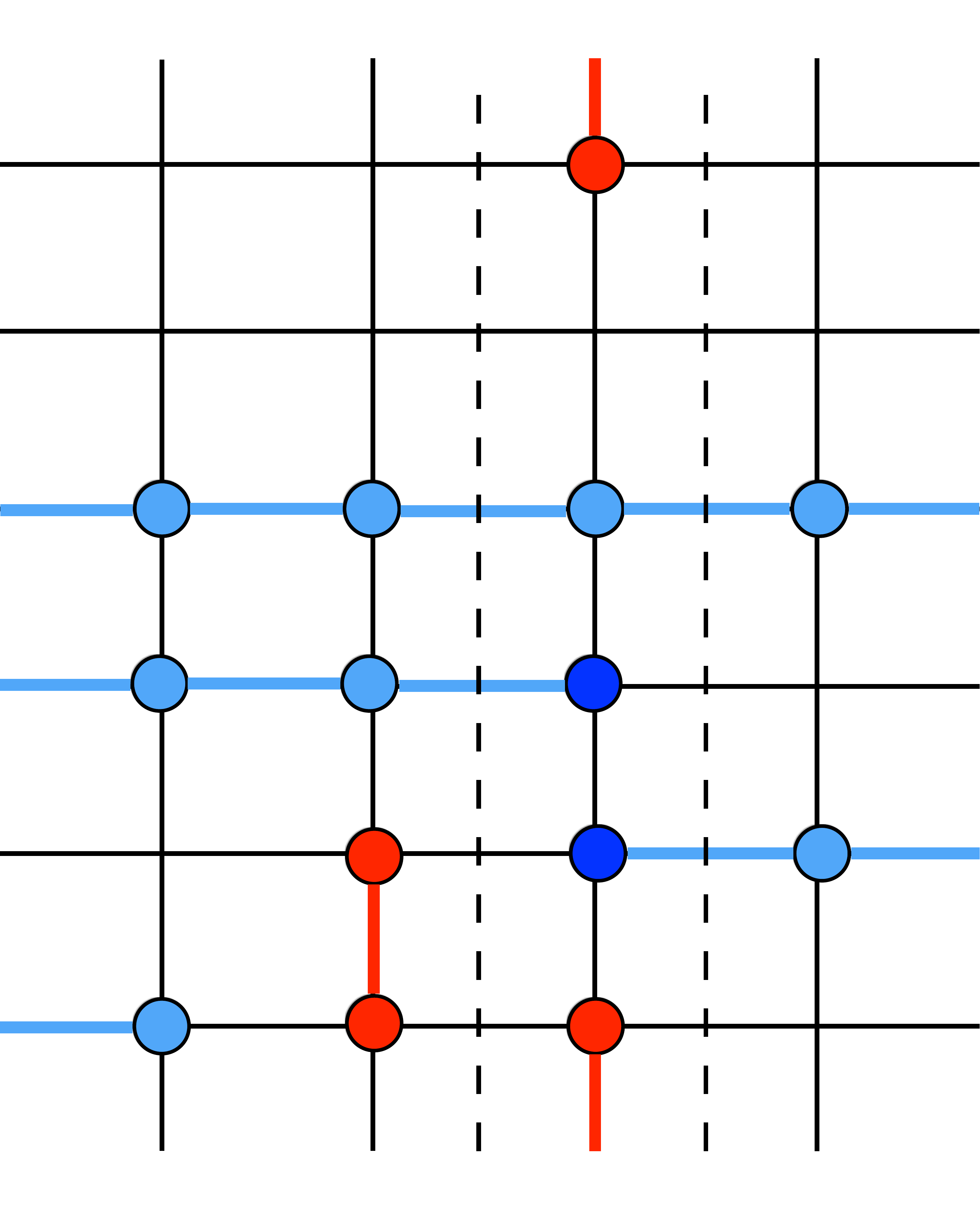}
\caption{Part of the $6\times L$ lattice. The dashed lines cross the edges which define the states of the transfer matrix.}
\label{6_L}
\end{figure}
\begin{figure}
\includegraphics[width=0.65\columnwidth]{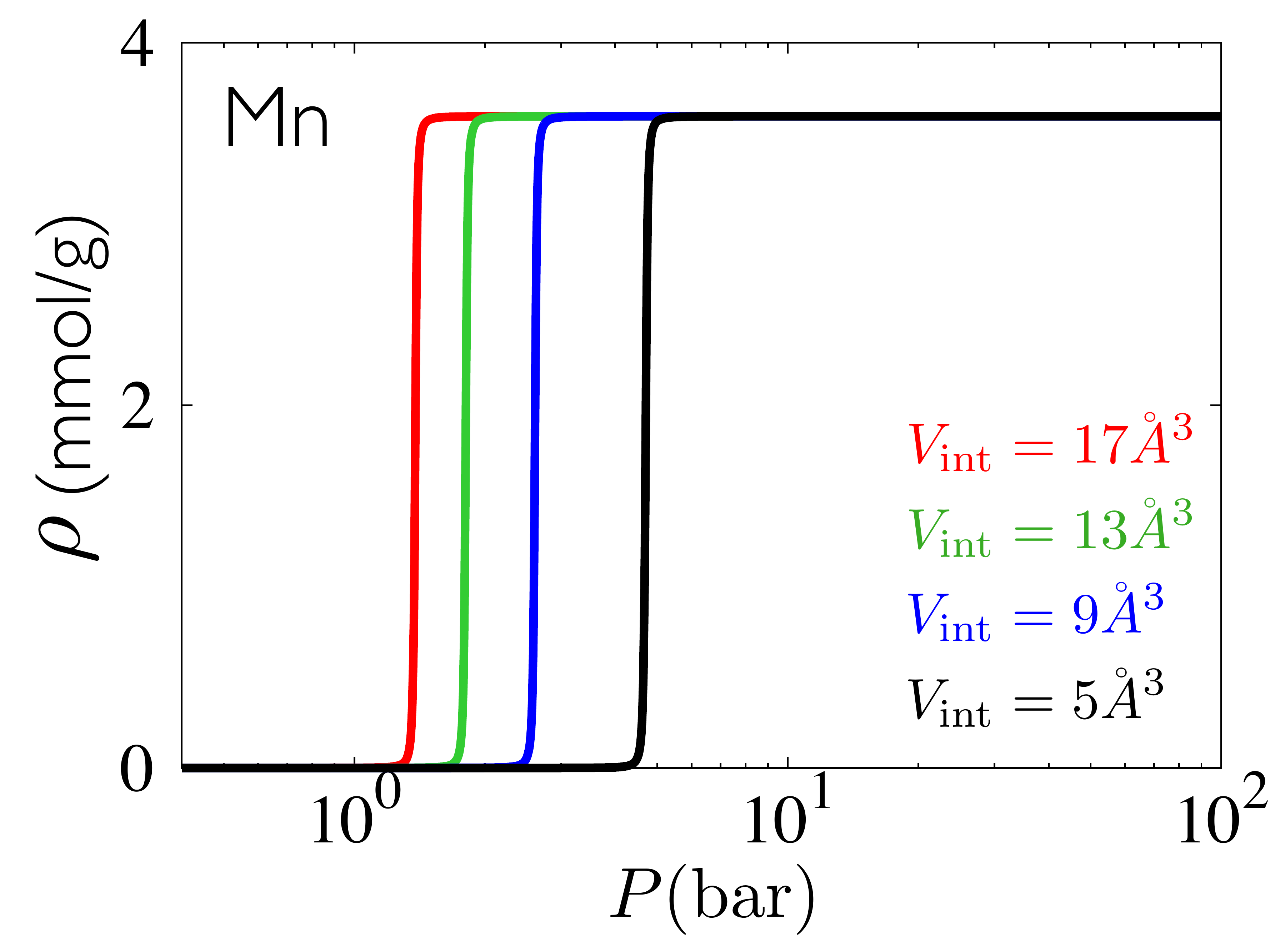}
\caption{Dependence of the step-position on $V_{\rm int}$: Isotherms obtained from the 1-lane model for mmen-$\mbox{Mn}_2$(dobpdc) at $313$ K for different values of $V_{\rm{int}}$. Here, $E_1$ and $E_{\rm{int}}$ are fixed to our calculated values (Table~\ref{fig_table}).}
\label{vint}
\end{figure}
\begin{figure}
\includegraphics[width=0.65\columnwidth]{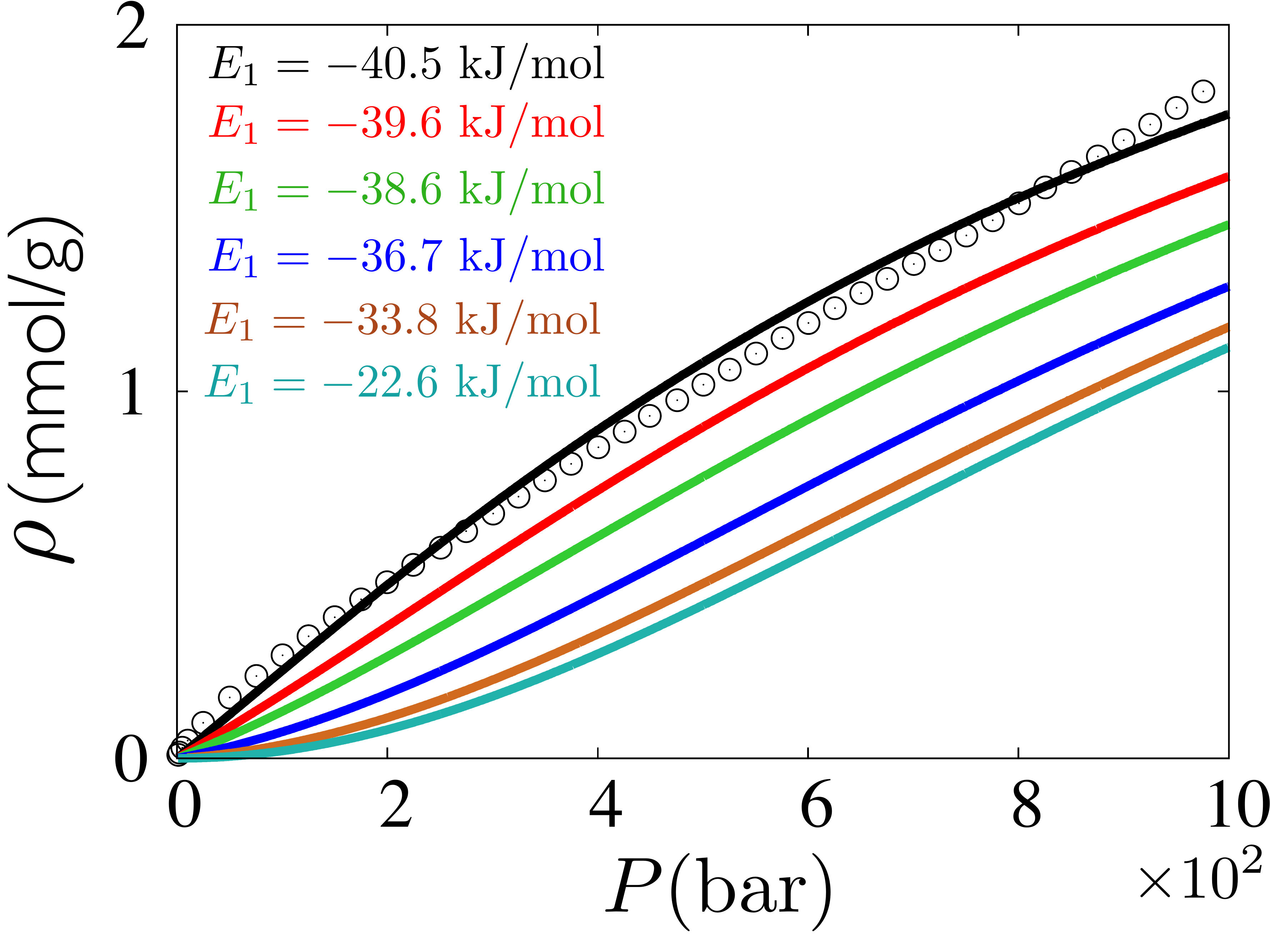}
\caption{Inflection point in the adsorption isotherm of mmen-Ni$_2$(dobpdc) disappears bellow a certain value of $E_1$: Isotherms for mmen-$\rm{Ni}_2$(dobpdc) when the binding energy of the single-bound \cot~conformation is varied at $313$ K. Theres exists a inflection point in the isotherm when $|E_1| \lesssim 40.5$ kJ/mol. Here, $E_{\rm int}=-46.4$ kJ/eV  ($V_{\rm init}=11 \AA^3$), $E_{\rm d}=-44.9$ kJ/eV, and $E_{\rm sec}=-42.0$ kJ/eV.}
\label{inflection}
\end{figure}
\section{The 2-lane model and the 6-lane model}
\label{2_6lane}
First, we start with the construction of the transfer matrix for the 2-lane system. Denote left and right lanes by $L$ and $R$. Define the restricted partition function $Z_N^{L,s_1;R,s_2}$ for two lanes of length $N$ with an external bond in each lane, specified to be $s_1,~s_2= u~\mbox{(unpolymerized) or}~p~\mbox{(polymerized)}$. Then we can write,
\be
\label{t2}
\left( \begin{array}{c} Z^{L,u;R,u}_{N+1} \\ \\ Z^{L,u;R,p}_{N+1} \\ \\ Z^{L,p;R,u}_{N+1} \\ \\ Z^{L,p;R,p}_{N+1}\end{array} \right) = \begin{pmatrix} 1+2 K_1+K_1^2+K_{\rm d}^2 && K_{\rm end}(1+K_1) &&  K_{\rm end}(1+K_1) && K_{\rm end}^2\\ \\ K_{\rm end}(1+K_1) && K_{\rm int}(1+K_1) && K_{\rm end}^2 && K_{\rm int} K_{\rm end} \\  \\ K_{\rm end}(1+ K_1) && K_{\rm end}^2 && K_{\rm int} (1+K_1) && g^2 K_{\rm int} K_{\rm end} \\  \\ K_{\rm end}^2 && K_{\rm int} K_{\rm end} &&K_{\rm int} K_{\rm end} && K_{\rm int}^2 \end{pmatrix} \left( \begin{array}{c} Z^{L,u;R,u}_{N} \\ \\ Z^{L,u;R,p}_{N} \\ \\ Z^{L,p;R,u}_{N} \\ \\ Z^{L,p;R,p}_{N} \end{array} \right),
\ee
where the statistical weights of a \cot~molecule in the single-bound conformation, in the pair (or dimer) conformation, as a part of chain-interior and chain end-points are given by $K_1$, $K_{\rm d}$, $K_{\rm int}$, and $K_{\rm end}$ respectively. In the thermodynamic limit the free energy is given by the logarithm of the largest eigenvalue of the matrix shown in \eqq{t2}. In this case, we can solve the largest eigen value and hence the free energy exactly. From the free energy, we calculate all the important thermodynamic quantities. 

For the 6-lane model with periodic boundary condition along the transverse direction (which behaves in all important respects like the 2-lane model) the calculation proceeds in an analogous way, yielding a $2^6 \times 2^6$-dimensional transfer matrix. Here we explain the elements of the transfer matrix. Chains are placed on the edges along the horizontal direction and pairs or dimers have transverse orientation, thus connecting two adjacent lanes (see \f{6_L}). The states which define the transfer matrix are given by the configurations of sets of horizontal edges, such as the ones crossed by the dashed lines in \f{6_L}, which may be empty or occupied by a bond of a chain. In the figure, the state of the left set of horizontal edges crossed by the first dashed line is $<i|=(0,0,1,1,0,0)$, while the corresponding state of the next set of horizontal edges crossed by the second dashed line is $<i+1|=(0,1,0,1,0,0)$. Here, 0 denotes an edge that is not a part of a chain, 1 denotes that the edge is a part of a chain. The contribution to the transfer matrix element ($T_{i,i+1}$) due to this particular pair or dimer configuration (one pair) is $(1 + K_1) K^2_{\rm d} K_{\rm int} K^2_{\rm end}$, since the remaining site may either be empty or occupied by a single-\cot. Thus, each element of the transfer matrix is a polynomial in the statistical weights. The particular element we are considering ($T_{i,i+1}$) has another contribution, with no dimer, which is $(1+K_1)^3 K_{\rm int} K_{\rm end}^2$. Thus, $<i|T|i+1>=T_{i,i+1}=2(1 + K_1) K^2_{\rm d} K_{\rm int} K^2_{\rm end}+(1+K_1)^3 K_{\rm int} K_{\rm end}^2$ (total six sites-- three of them are occupied by chains; the remaining three sites we can either have one pair and one single-\cot~or no pairs and three single-\cot~molecules; the factor of $2$ is the multiplicity of the configuration with a single pair). In a compact form, the contribution from a configuration with $j$ number of pairs, $k$ number of chain-interior sites, and $\ell$ number of chain end-points is $i (1+K_1)^{6-2j-k-\ell} K_{\rm d}^{2j} K^k_{\rm int} K^{\ell}_{\rm end}$, where $i$ is the multiplicity, and $(6-2j-k-\ell)$ is the number of remaining sites that may either be empty or occupied by single-\cot~molecules. Hence, there are total $2^6$ possible states--$(\eta_1,\eta_2,\dots,\eta_6)$ where $\eta_i=0~\rm{or}~1$, so that $<1|=(0,0,0,0,0,0),~<2|=(0,0,0,0,0,1),~<3|=(0,0,0,0,1,0),\dots,<64|=(1,1,1,1,1,1)$. This leads to a transfer matrix of size $2^6 \times 2^6$. One can compute all the matrix elements exactly, and calculate the largest eigenvalue numerically. For example, $T_{1,1}=(1+K_1)^6+6(1+K_1)^4 K^2_{\rm d}+9(1+K_1)^2 K^4_{\rm d}+2 K^3_{\rm d}$. 

\section{Estimation of free volumes}
\label{free_vol}
As discussed in the main text, \cot~can be adsorbed in three different conformations-- as a single-bound molecule, as a part of a bound (carbamic acid) pair, or as a part of a (ammonium carbamate) chain. Here, we estimate the free volume accessible to a \cot~molecule in each conformation using very crude geometric arguments. In the single-bound conformation, the ligand along with the physisorbed \cot~molecule possesses orientational and conformational degrees of freedom. The ligand with a bound \cot~can roughly access a free volume of a hemispherical shell with radius $\sim 2.3-6.3 \AA$ (typical euclidean distance between the metal site and \cot~considering various conformations). Thus we set, $V_1=\frac{1}{2}\frac{4}{3} \pi (6.3^3-2.3^3)\approx 500 \AA^3$. In the pair conformation, each \cot~molecule along with the ligand can access roughly the volume of half of a ring with radius $5.5-6.5 \AA$ (typical distance of the pair from the surface of the framework) and width $\sim 4 \AA$ (wiggle room of \cot~molecules with in the pair conformation). We set, $V_{\rm d}=\frac{1}{2} \pi (6.5^2-5.5^2)\times 4 \approx 75 \AA^3$. The chain conformation is almost frozen and allows a very small free volume. A chemisorbed \cot~molecule within a chain can access roughly the volume of half of a ring with radius $1.8-2.8 \AA$ and width $1.5 \AA$ (wiggle room of a \cot~molecule with in a chain). We set $V_{\rm{int}}=V_{\rm{end}}=\frac{1}{2} \pi (2.8^2-1.8^2)\times 1.5 \approx 11 \AA^3$. It is worth noting that, for a step-like isotherm, $V_{\rm{int}}$ plays the most significant role in determining the step-position (see \f{vint}). The values of $V_1$ and $V_{\rm d}$ only affect the rise of the isotherm at low pressure before the step. Their effect is negligible if the chain conformation is energetically much more favorable. In the case of physisorbed \cot~molecules at the secondary binding sites, each molecule can access roughly the volume of a hemispherical shell with radius $\sim 1.9-3.8 \AA$ (distance between the binding site and \cot). Thus, we choose, $V_{\rm{sec}}=\frac{1}{2}\frac{4}{3} \pi (3.8^3-1.9^3)\approx 100 \AA^3$. These measurements do not take into account all the possible conformations of the ligands, and likely underestimate the conformational entropy, but overestimate the free volume, due to the neglect of strong steric effects present in the real system. Deviations from these estimated values do not change the qualitative feature of the isotherms.

\section{Capturing fine details of isotherms}
\label{details}

In the main text we focus on describing the basic physics of cooperative binding, which is captured by a simple one-lane statistical mechanical model. Here we show, in addition, that fine features of isotherms, such as the rise before and after the sharp step can be captured by considering the 2- or 6- lane models. For some metals, in addition, we show that reproduction of these features requires inclusion of secondary binding sites and a different mode of monomer binding within the model.

For mmen-M$_2$(dobpdc) built with the metals M = Fe, Co, or Zn, the bound-pair conformation is energetically significant, and thus we solve the 6-lane (and the 2-lane) system by generalizing the transfer matrix formalism presented in the main text. Although the basic physics of cooperative adsorption in these frameworks is captured by the 1-lane model, which ignores the pair conformation, the experimental system is more closely represented by the 2- or 6-lane model (the 1-lane model ignores the pair conformation).

Here, we introduce statistical weights of a \cot~in the pair (dimer) conformation, $K_{\rm d}=g_{\rm d} W_{\rm d}=g_{\rm d} e^{-\beta (E_{\rm d}-\mu)}$. $E_{\rm d}$ represents the binding energy of \cot~in the pair conformation, and $g_{\rm d}=V_{\rm d}/\Lambda^3$, where $V_{\rm d} (\approx 75 \AA^3)$ is the free volume accessible to a \cot~molecule in that conformation. For an $n$-lane system, the transfer matrix is of dimension $2^n\times 2^n$, and the largest eigenvalue can be solved exactly or numerically (see Sec.~\ref{2_6lane}). The 6-lane model behaves in all important respects like the 2-lane model.
\begin{figure}
\includegraphics[width=0.65\columnwidth]{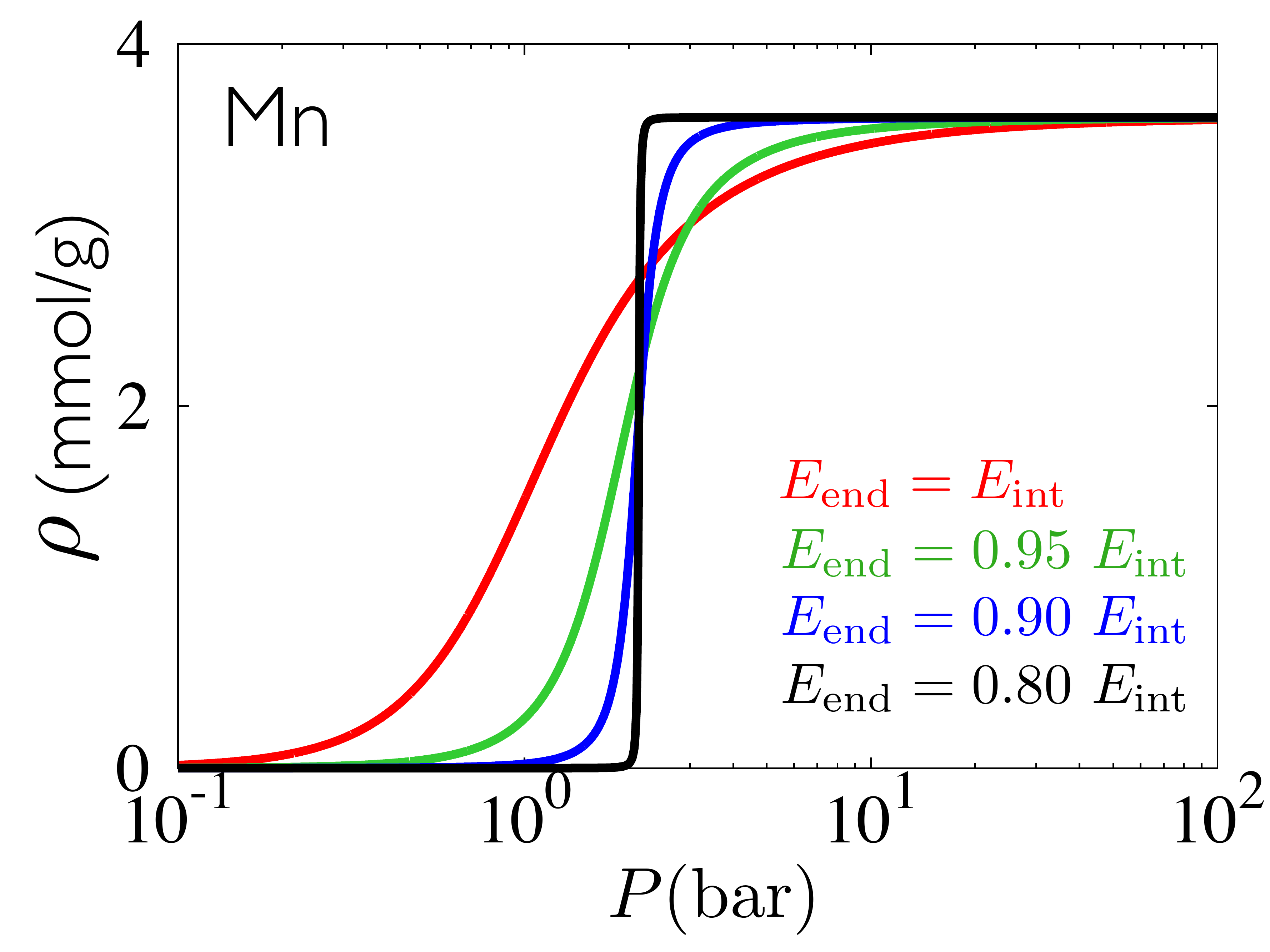}
\caption{Variation of the slope of a isotherm as a function of $E_{\rm end}$: isotherms obtained from the 1-lane model for mmen-$\mbox{Mn}_2$(dobpdc) at $313$ K for different chosen values of the binding enthalpy of chain end-points, where $E_1$ and $E_{\rm{int}}$ are fixed to our calculated values (Table~\ref{fig_table}) and $V_{\rm{int}}=11\AA^3$.}
\label{endpoints}
\end{figure}

Isotherms derived from the 6-lane model are shown in \f{isotherm} (c--f) in the main text, and match key features (shape, scaling of the step-position with temperature) of the experimental data. The step-shaped isotherms are similar to those of the 1-lane model, with additional corrections at low pressure (before the step) that results from pair-binding (noticeable in case of Co: \f{isotherm} d left panel in the main text). The 1-lane model fails to capture this feature when the single-\cot~molecules have significantly low binding affinity. Within the current DFT-based energetics, we do not see a prominent slow rise of the isotherm before the step feature for Fe, and Zn in contrast to the experimental data.
We find that the step-position is sensitive to statistical weights of chain-interior sites (see \f{vint}) and we cannot get quantitative agreement with the experiments using a unique value of $V_{\rm{int}}$ given the energetics. We shall come back to this issue in the next paragraphs. 

The case of Ni is slightly more subtle. Using the DFT energetics shown in Table~\ref{fig_table} we find that the Ni-based framework shows no sharp step in its isotherm, reproducing the experimental observation\c{david2015}. Here the chain conformation is no longer statistically the most favorable one.
However, we see an inflection point at low pressure in the isotherm for Ni that seems to be absent in experiment (see~\f{isotherm} f in the main text). The microscopic origin of this inflection point is the very low probability of the single-bound molecules compared to the pairs. The experimental heat of adsorption curve for Ni exhibits a very different feature than the other metals-- it shows a single plateau at $\approx -40$ kJ/mol at high loading\c{david2015}. Interestingly, we find that only with a larger binding affinity for the single-\cot~conformation ($\approx -40.5$ kJ/mol) does the inflection point disappear \f{inflection}. Here, we propose that there might exist some other single-bound species (e.g. chains of length one) that are energetically comparable with pairs. This does not have any noticeable effect on the position or slope of the step if it exists. Alternatively, the step-like feature may also disappear if the statistical weight of the adsorbate~at the chain-interior and at the chain end-points are similar (see \f{1d_isotherm} d). 

Our model can be used to determine how to use microscopic parameters to control gas-uptake isotherms. For instance, one can enhance cooperativity by making the chain end-points less favorable as this leads to a sharper step (see \f{endpoints}). The isotherm for mmen-Ni$_2$ (dobpdc) shown in \f{isotherm} f (in the main text) is non-cooperative. It can be made cooperative by enhancing the effective binding affinity of \cot~within a chain, as shown in \f{multi}. We check this by scaling the statistical weights for chain binding, $K_{\rm end}$ and $K_{\rm int}$, by a factor $1+K_{\rm sec}$, leading to energetic stabilization of chains through an additional species that doesn't compete with \cot~for primary binding sites. We can write $K_{\rm sec}= g_{\rm sec} e^{-\beta (\Delta E-\mu)}$, where $g_{\rm sec}=V_{\rm sec}/\Lambda^3$ ($V_{\rm sec}$ is the free volume accessible to that additional species in the bound state) and predict the required binding-affinity enhancement $\Delta E$ to induce cooperativity (i.e., a step) at a suitable pressure. There may be several possible ways of engineering such enhancement. For instance, mmen-M$_2$(dobpdc) is known to exhibit {\em enhanced} \cot~uptake in the presence of \water\c{long_jacs2015}, which is striking as water competes with \cot~for binding sites in most MOFs\c{long_jacs2015,piero2013,kundu2016}. This factor of $1+K_{\rm sec}$ can also be regarded as a way of modeling the existence of adsorption at secondary binding sites. In experiments (see \f{isotherm} in the main text), a slow rise of the isotherms after the step is sometimes observed with increasing pressure, which indicates the existence of secondary binding sites. 

Finally, considering a different mode of monomer binding as discussed and adsorption at the secondary binding sites, we can reproduce all the features of the experimental isotherms: the rise of the isotherms at low pressures (even in the case of Fe, Zn); the step; followed by the rise of the isotherm at high pressures due to physisiorption of \cot~at secondary binding sites (we set  the \cot-binding energy at the secondary binding sites as $E_{{\rm sec}} \approx -42.0$ kJ/mol, and the corresponding free volume $V_{\rm{sec}}\approx 100 \AA^3$)-- see \f{isotherm} (in the main text) right panels for all metals. Here, we use the experimental $E_{\rm{int}}$ values and set the values of $V_{\rm{int}}$ such that the step-position (if it exists) matches with that of the experimental isotherms (see Tables~\ref{fig_table} and~\ref{table02}). The discrepancy of the maximum uptake (after the step) may result from the defects in the experimental system where all the primary binding sites are  inaccessible to \cot\c{queen2014}. For Ni, we use the DFT-based binding affinity (as the experimental data is not available) for the chain conformation; tune the pair binding energy and set $E_{{\rm d}} \approx -44.9$ kJ/mol as the best-fit value.

In conclusion, we attribute the rise of the isotherm at low pressures to the pairs or the predicted single-\cot~conformation; the existence of the step to the formation of long chains (the step position is controlled by the binding energy of chain-interior sites); and the slow rise of the isotherm at very high pressures to adsorption at secondary binding sites. There exists no direct proof of the existence of the pair conformation. Even without the pairs, however, the existence of a possible different mode of single-\cot~binding is sufficient to reproduce all the fine features of the experimental isotherms with the $1$-lane system (see \f{1d_isotherm}). The existence of this predicted new mode of monomer binding can be tested with extensive DFT calculations.
\begin{figure}
\includegraphics[width=0.85\columnwidth]{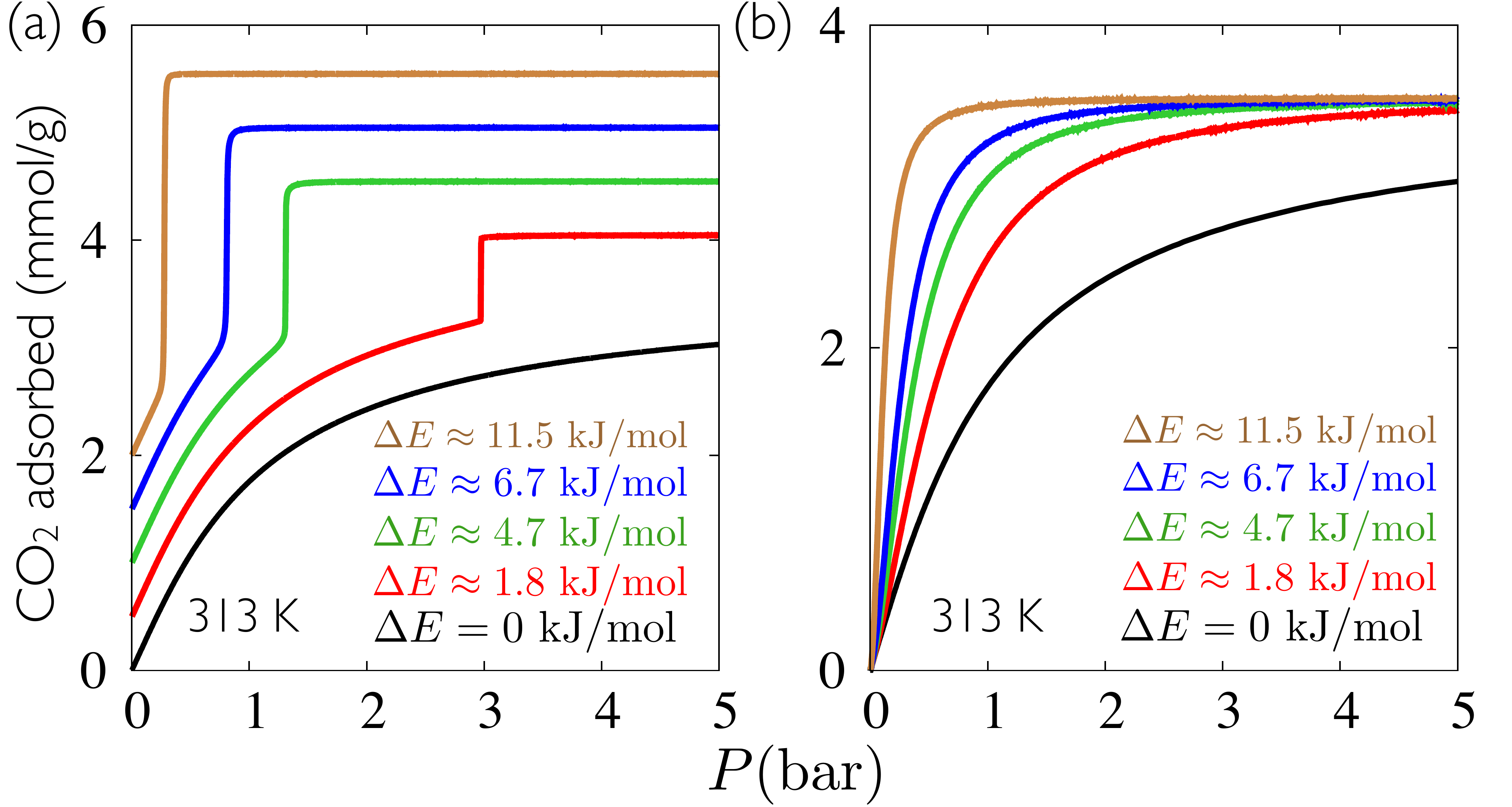}
\caption{Induced cooperativity by an agent species: (a) Isotherms for our (6-lane) model of mmen-Ni$_2$(dobpdc) in the presence of an agent that doesn't compete with \cot~but enhances the binding affinity of a polymerized \cot~molecule. This is achieved by multiplying a factor $(1+K_{\rm sec})$ to $K_{\rm int}$ and $K_{\rm {end}}$. For a binding enhancement $\Delta E$ exceeding about $\sim 6$ kJ/mol, we predict that the non-cooperative Ni isotherm can be made cooperative around $1$ bar. For clarity, curves have been shifted along the $y$-axis by increments of 0.5, 1.0, 1.5 and 2.0 mmol/gm, respectively. (b) Isotherms in the presence of an agent that doesn't compete with \cot~but enhances the binding affinity of \cot~in all the three conformations- single, pair, and chain. This is achieved by multiplying a factor $(1+K_{\rm sec})$ uniformly with $K_{\rm int}$, $K_{\rm {end}}$, $K_{\rm d}$ and $K_1$. In this case, the step-like feature doesn't appear.}
\label{multi}
\end{figure}
\begin{figure}
\includegraphics[width=0.7\columnwidth]{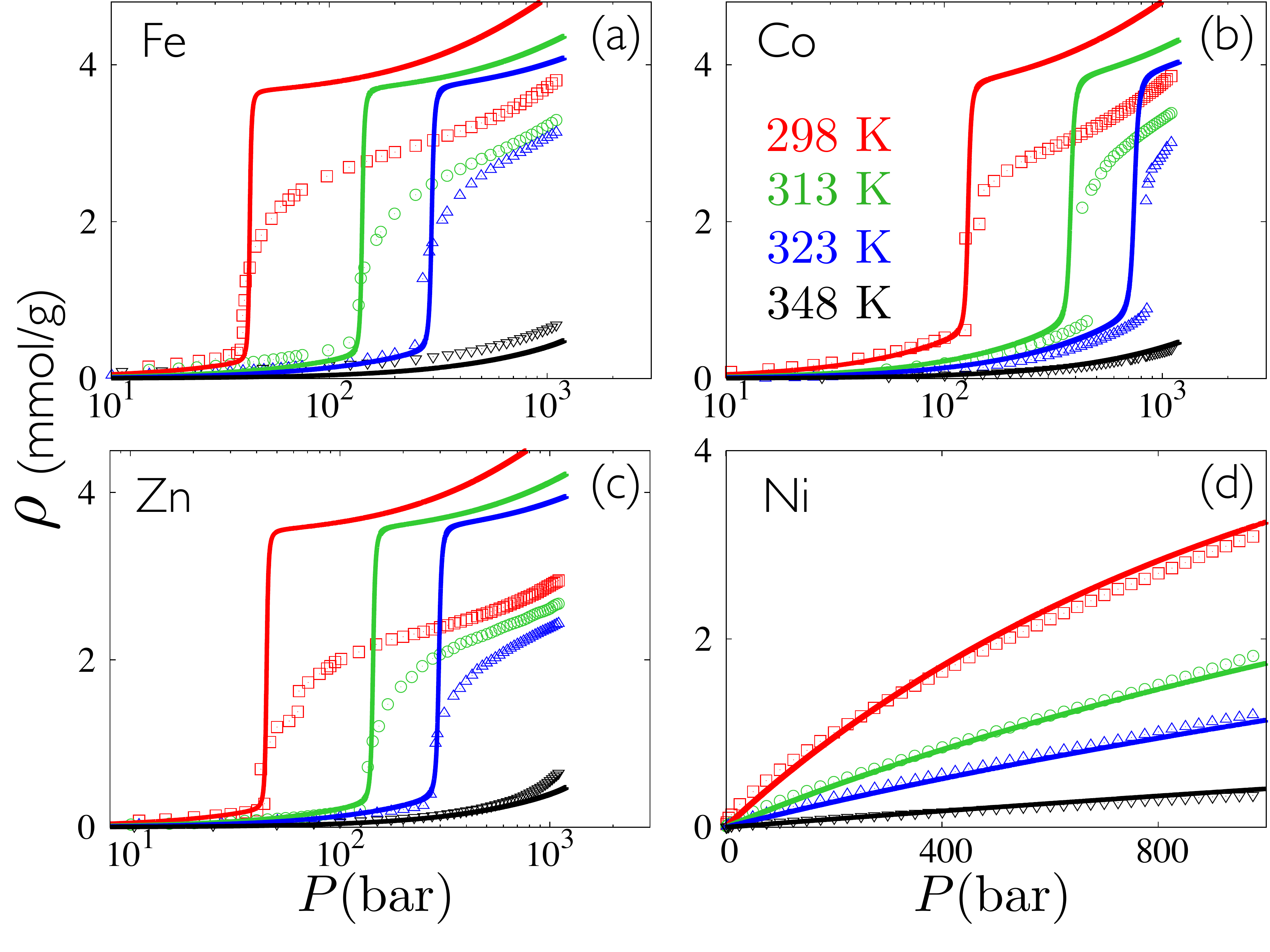}
\caption{Isotherms (lines) derived from the 1-lane model of mmen-$\mbox{M}_2$(dobpdc), where M = Fe, CO, Zn, or Ni, agrees well with the experiments. The binding enthalpies of the chain conformation for different metals are set by the experimental values (see Table~\ref{fig_table}).  For Fe, Co, and Zn the values of $V_{\rm{int}}$ are taken from Table~\ref{table02}. For Ni, we set $E_{\rm{int}}=E_{\rm end}=-46.4$ kJ/mol (see Table.~\ref{fig_table}) and $V_{\rm{int}}=23 \AA^3$ to get the best fit. Here, we take into account the predicted new mode of single-\cot~binding and adsorption at the secondary binding sites. Data points are from experiments\c{david2015}.}
\label{1d_isotherm}
\end{figure}

\section{Chain-length distribution for the 1-lane model}
\label{cld}

\begin{figure}[b]
\includegraphics[width=0.7\linewidth]{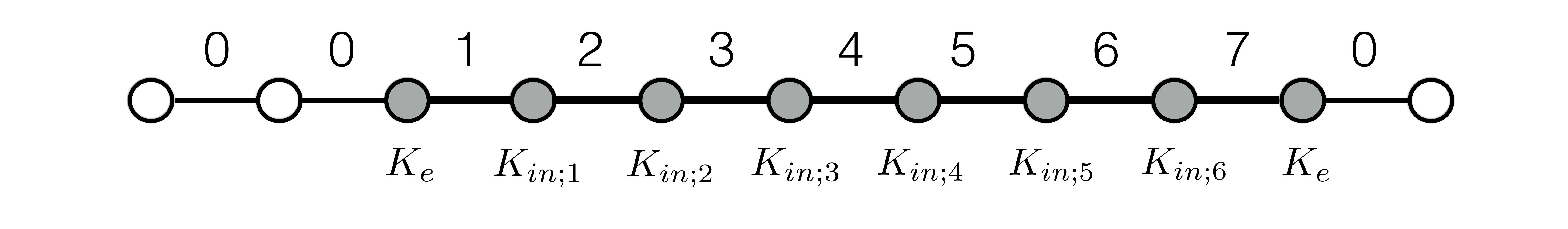}
\caption{Construction used to calculate the chain-length distribution for the 1-lane model. The figure shows a chain of length $8$ with two end-points and $6$ internal monomers. The numbers over the bonds are the variables $\eta$, mentioned in the text}
\label{chain_config}
\end{figure}
The grand partition function of the system is given by
\bea
\mathcal{Z}=\sum_{\{n_1,n_{\rm int},n_{\rm end}\}} K_1^{n_1} K_{\rm end}^{n_{\rm end}} K_{\rm int}^{n_{\rm int}} \Gamma (n_1,n_{\rm int},n_{\rm end}),
\eea
where
\bea
\Gamma(n_1,n_{\rm int},n_{\rm end}) &=&  \frac{(N-n_{\rm int}-n_{\rm end}/2)!}{(N-n_1-n_{\rm int}-n_{\rm end})! (n_{\rm end}/2)!n_1! }  \frac{(n_{\rm end}/2+n_{\rm int}-1)!}{(n_{\rm end}/2-1)! n_{\rm int}!}
\label{eq_gamma}
\eea
is the number of ways of arranging $n_1$ single \cot~molecules, $n_{\rm int}$ internal chain molecules and $n_{\rm end}$ chain end-points on a $1d$ lattice with $N$ sites. Here, we do not distinguish between different chain-interior sites. 

 To determine the chain-length distribution for the 1-lane model, we introduce position-dependent statistical weights for internal chain monomers, as shown in \f{chain_config}. The weights of the internal chain monomers starting from the left-hand side are denoted $K_{{\rm int};1}$, $K_{{\rm int};2}$, etc. The terminal chain monomers have weight $K_{\rm end}$. Bonds internal to the chain are denoted $\eta=1,2,\dots$, counting from the left. For bonds not occupied by chains we set $\eta=0$. To write down the transfer matrix for this case, we again define the restricted partition functions $Z^{\eta}_N$. This is the partition function for a system of $N$ sites, with an edge added to the outside of the $N^{\rm th}$ site with bond variable $\eta$. 

The restricted partition functions satisfy
\be
 \left( \begin{array}{c} Z^0_{N+1} \\ \\ Z^1_{N+1} \\ \\ Z^3_{N+1} \\ \\ Z^4_{N+1} \\ \\  \vdots \end{array} \right) = \begin{pmatrix} 1+K_1 && K_{\rm end} &&  0 && 0 && \dots\\ \\ K_{\rm end} && 0 && K_{{\rm int};1} && 0 && \dots \\  \\ K_{\rm end} && 0 && 0 && K_{{\rm int};2} && \dots \\  \\K_{\rm end} && 0 && 0 && 0 && \dots \\ \\  \vdots  &&  \vdots &&  \vdots &&  \vdots &&  \vdots \end{pmatrix} \left( \begin{array}{c} Z^0_N \\ \\ Z^1_N \\ \\ Z^3_N \\ \\ Z^4_N \\ \\  \vdots  \end{array} \right),
\ee
where $T$ is the matrix shown. The secular equation can be obtained by calculating the determinant $|T-\lambda I|$, and is
\be
-1+\frac{1}{\lambda}+\frac{K_1}{\lambda}+\left(\frac{K_{\rm end}}{\lambda}\right)^2 \left( 1+\sum_{i=1}^{\infty}\frac{1}{\lambda^i}\displaystyle\prod_{j=1}^{i} K_{{\rm int};j}\right)=0.
\label{secular}
\ee
The density of internal monomers in the $m^{\rm th}$ position is
\be
\rho^{{\rm ch}}_m=\frac{K_{{\rm int};m}}{\lambda} \frac{\partial \lambda}{\partial K_{{\rm int};m}}.
\ee
Differentiating both sides of \eqq{secular} and setting $K_{{\rm int};1} =K_{{\rm int};2}=\dots = K_{\rm int}$ gives
\be
\rho^{{\rm ch}}_m=\frac{K_{\rm end}^2 \omega (1-\omega K_{\rm int})}{(1+K_1)(1-\omega K_{\rm int})^2+K_{\rm end}^2 \omega (2-\omega K_{\rm int})} (\omega K_{\rm int})^m,
\label{rho_m}
\ee
where $\omega\equiv 1/\lambda_+$ (see \eqq{ev}). When all internal monomers are equivalent, the densities of end-point chain monomers ($\rho_{\rm end}$), internal chain monomers ($\rho_{\rm int}$) and single bound \cot~molecules ($\rho_1$) may be obtained taking the derivative of \eqq{secular} with respect to  $K_{\rm end}$, $K_{\rm int}$ and $K_1$ respectively. Doing so gives
\bea
\rho_{\rm end}=\frac{2 K_{\rm end}^2 \omega (1-\omega K_{\rm int})}{(1+K_1)(1-\omega K_{\rm int})^2+K_{\rm end}^2 \omega (2-\omega K_{\rm int})}, \label{rho_e} \\
\rho_{\rm int}=\frac{K_{\rm end}^2 \omega (\omega K_{\rm int})}{(1+K_1)(1-\omega K_{\rm int})^2+K_{\rm end}^2 \omega (2-\omega K_{\rm int})}, \label{rho_in} \\
\rho_1=\frac{K_1 (1-\omega K_{\rm int})^2}{(1+K_1)(1-\omega K_{\rm int})^2+K_{\rm end}^2 \omega (2-\omega K_{\rm int})}.
\label{rho_1}
\eea
Using \eq{rho_e} and~\eq{rho_in}, one may rewrite \eq{rho_m} in terms of the densities of terminal and internal chain monomers as 
\be
\rho^{{\rm ch}}_m=\frac{\rho_{\rm end}}{2} \left(\frac{2 \rho_{\rm int}}{\rho_{\rm end}+2\rho_{\rm int}}\right)^m.
\ee
Now we can compute the fraction of chains of length $\ell$ ($\ell-2$ internal monomers):
\be
\begin{split}
r_\ell&=\frac{2}{\rho_{\rm end}} \left(\rho^{{\rm ch}}_{\ell-2}-\rho^{{\rm ch}}_{\ell-1}\right) \\
&=(\omega K_{\rm int})^{\ell-2} (1-\omega K_{\rm int}) \\
&=\left(\frac{K_{\rm int}}{\lambda_+}\right)^{\ell-2} \left(1-\frac{K_{\rm int}}{\lambda_+}\right) \\
&=\frac{(2 K_{\rm int})^{\ell-2}\left[1+K_1-K_{\rm int}+\sqrt{(1+K_1-K_{\rm int})^2+4 K_{\rm end}^2}\right]}{\left[1+K_1+K_{\rm int}+\sqrt{(1+K_1-K_{\rm int})^2+4 K_{\rm end}^2}\right]^{\ell-1}}.
\end{split}
\label{dist_m}
\ee
We may also rewrite \eqq{dist_m} in terms of $\rho_{\rm end}$ and $\rho_{\rm int}$ as 
\be
r_\ell=\frac{\rho_{\rm end}}{\rho_{\rm end}+2 \rho_{\rm int}}\left(\frac{2 \rho_{\rm int}}{\rho_{\rm end}+2 \rho_{\rm int}}\right)^{\ell-2}.
\ee
In the limit $P \to \infty$ the mean chain length approaches the asymptotic value 
\bea
\label{li}
\av{\ell}_{\infty}=\frac{2 V_{\rm int} e^{-\beta E_{\rm int}}} {V_1 e^{-\beta E_1} -V_{\rm int} e^{-\beta E_{\rm int}}+\sqrt{(V_1 e^{-\beta E_1} -V_{\rm int} e^{-\beta E_{\rm int}})^2+4V_{\rm int}^2 e^{-2\beta E_{\rm end}}}}+2;
\eea
note that $K_\alpha=\beta P V_\alpha e^{-\beta E_{\alpha}}$.

\section{Correlations in $1 \times L$ model}
\label{s1}
The transfer matrix for a single lane is
\begin{equation}
T=\begin{pmatrix}1+K_1&K_{\rm end}\\K_{\rm end}&K_{\rm int}\end{pmatrix}.
\end{equation}
When the weight of endpoint monomers $K_{\rm end}$ vanishes, the matrix is diagonal and the eigenvalues are $\lambda_1=1+K_1$ and $\lambda_2=K_{\rm int}$. The eigenvectors are:
\begin{eqnarray}
\phi_1&=&\begin{pmatrix}1\\0\end{pmatrix},\\
\phi_2&=&\begin{pmatrix}0\\1\end{pmatrix},
\end{eqnarray}
respectively. If the lattice edge $i$ is occupied by a bond, the state variable $\eta_i=1$, otherwise we have $\eta_i=0$. We will consider periodic boundary conditions here. We may then define the expectation values $\langle \eta_i \rangle$ and $\langle \eta_i \eta_j \rangle$, where we assume $j>i$. The bond-bond correlation function will the be $c_{i,j}=\langle \eta_i\eta_j\rangle-\langle \eta_i \rangle \langle \eta_j \rangle$. Since we have translational symmetry, $\langle \eta_i \rangle=\langle \eta \rangle$, independent of $i$, and $c_{i,j}$ is a function of the distance $j-i$. For vanishing $K_{\rm end}$, we have $\langle \eta \rangle=0$ if $1+K_1>K_{\rm int}$ and $\langle \eta \rangle=1$ if $1+K_1<K_{\rm int}$, while $c_{i,j}$ vanishes identically.

When the weight of endpoint monomers does not vanish, the eigenvalues are $\lambda_1=(K+2K_{\rm int}+\sqrt{K^2+4K_{\rm end}^2})/2$ and $\lambda_2=(K+2K_{\rm int}-\sqrt{K^2+4K_{\rm end}^2})/2$, where $K=1+K_1-K_{\rm int}$. The eigenvectors are
\begin{equation}
\phi_{1,2}=\frac{1}{\sqrt{1+a_{1,2}^2}}\begin{pmatrix}1\\a_{1,2}\end{pmatrix},
\end{equation}
where $a_1=2K_{\rm end}/\left(K+\sqrt{K^2+4K_{\rm end}^2}\right)$ and $a_2=2K_{\rm end}/\left(K-\sqrt{K^2+4K_{\rm end}^2}\right)$. The partition function for a lattice with $N$ sites is:
\begin{equation}
Z_N=\sum_{\{\eta\}}\prod_{i=1}^NT(\eta_i,\eta_{i+1})=\sum_{\eta_1=0,1}T^N(\eta_1,\eta_1)=\lambda_1^N + \lambda_2^N,
\end{equation}
where $\eta_{N+1}=\eta_1$ because of the boundary conditions. The density of bonds per site is:
\begin{equation}
\langle \eta \rangle=\frac{1}{Z_N}\sum_{\eta_1=0,1}\eta_1T^N(\eta_1,\eta_1)=\frac{1}{Z_N} T^N(1,1).
\end{equation}
Now, we have:
\begin{equation}
T^s(\eta,\eta^\prime)=\sum_{i=1,2}\lambda_i^s\phi_i(\eta)\phi_i(\eta^\prime),
\end{equation}
so that:
\begin{equation}
\langle \eta \rangle=\frac{1}{\lambda_1^N+\lambda_2^N}\sum_{i=1,2}\lambda_i\frac{a_i^2}{1+a_i^2},
\end{equation}
and
\begin{eqnarray}
\langle\eta_1 \eta_k \rangle&=&\frac{1}{Z_N}\sum_{\{\eta\}}\eta_1T(\eta_2,\eta_2)\ldots T(\eta_{k-1},\eta_k)\eta_kT(\eta_k,\eta_{k+1})\ldots T(\eta_N\eta_1) = \nonumber \\
&&\frac{1}{Z_N}T^{k-1}(1,1)T^{N-k+1}(1,1)= \nonumber \\
&&\frac{1}{\lambda_1^N+\lambda_2^N} \left( \sum_{i=1,2} \lambda_i^{k-i}\frac{a_i^2}{1+a_i^2}\right)
\left( \sum_{i=1,2} \lambda_i^{N-k+1}\frac{a_i^2}{1+a_i^2}\right).
\end{eqnarray}
In the thermodynamic limit $N \to \infty$, since $\lambda_1>\lambda_2$, we have
\begin{equation}
\langle \eta \rangle=\frac{a_1^2}{1+a_1^2},
\end{equation}
and
\begin{equation}
\langle \eta_1\eta_k\rangle=\left(\frac{a_1^2}{1+a_1^2}\right)^2+\frac{a_1^2a_2^2}{(1+a_1^2) (1+a_2^2)}\left(\frac{\lambda_2}{\lambda_1}\right)^{k-1}.
\end{equation}
Therefore, the correlation function is:
\begin{equation}
c_{i,j}=\frac{a_1^2a_2^2}{(1+a_1^2) (1+a_2^2)}\exp \left(-\frac{j-i}{\xi}\right),
\end{equation}
where the correlation length is:
\begin{equation}
\xi=\frac{1}{\ln(\lambda_1/\lambda_2)}.
\end{equation}

\begin{center}
\begin{table}
\begin{tabular}{ |c|c|} 
 \hline
 M & \cot~capacity (mmol/g) \\ 
 \hline
 Mg & $4.040 $ \\ 
 \hline
 Mn & $3.595 $ \\ 
 \hline
 Fe & $3.583  $ \\ 
 \hline
 Co & $3.544 $ \\ 
 \hline
 Zn & $3.465 $ \\ 
 \hline
 Ni & $3.547 $ \\ 
 \hline
\end{tabular}
\caption {Maximum \cot~uptake capacity of mmen-$\mbox{Mg}_2$(dobpdc), assuming one \cot~per metal-diamne\c{david2015}.}
\label{table01}
\end{table}
\end{center}

\end{document}